\documentclass[prd,aps,nofootinbib,floatfix,12pt]{revtex4}
\usepackage{changes}
\usepackage{amsfonts,amsmath,amssymb}
\usepackage{graphicx,color,epsfig,epstopdf}
\usepackage{diagbox}
\usepackage{rotating}

\newcommand{\nn}{\nonumber}
\newcommand{\be}{\begin{equation}}
\newcommand{\ee}{\end{equation}}
\newcommand{\bea}{\begin{eqnarray}}
\newcommand{\eea}{\end{eqnarray}}
\newcommand{\ba}{\begin{array}}
\newcommand{\ea}{\end{array}}
\newcommand{\bi}{\begin{itemize}}
\newcommand{\ei}{\end{itemize}}

\newcommand{\mce}{{\mathcal E}}

\newcommand{\mch}{{\mathcal H}}

\newcommand{\difd}{\mathrm d}

\newcommand{\xbj}{x_B}

\usepackage{todonotes}
\newcounter{mycomment}

\newcommand{\ucas}{\affiliation{University of Chinese Academy of Sciences, Beijing 100049, China}}

\newcommand{\keylab}{\affiliation{State Key Laboratory of Heavy Ion Science and Technology, Institute of Modern Physics, Chinese Academy of Sciences, Lanzhou 730000, China}}

\newcommand{\hebei}{\affiliation{Department of Physics, Hebei University, Baoding 071002, China}}

\newcommand{\Zagreb}{\affiliation{Department of Physics, University of Zagreb Faculty of Science, HR-10000 Zagreb, Croatia}}

\begin{document}

\title{Assessing the impact of the electron ion collider in China \\ on Deeply Virtual Compton Scattering}

\author{Yuan-Yuan Huang}
\hebei
\keylab

\author{Xu Cao}\email{caoxu@impcas.ac.cn}
\keylab
\ucas

\author{Taifu Feng}\email{fengtf@hbu.edu.cn}
\hebei

\author{Kre\v{s}imir Kumeri\v{c}ki}\email{kkumer@phy.hr}
\Zagreb

\author{Yu Lu}\email{ylu@ucas.ac.cn}
\ucas


\date{\today}

\begin{abstract}
  \rule{0ex}{3ex}
    We assess the impact of future measurements of deeply virtual Compton scattering (DVCS) off protons using the planned detector at the Electron-Ion Collider in China (EicC), proposed as an upgrade to the High Intensity heavy-ion Accelerator Facility (HIAF). We develop a neural-network architecture to flexibly parameterize the Compton Form Factors (CFFs), extrapolate reliably into unmeasured kinematic regions, and provide robust uncertainty estimates through the replica method. The framework is fitted to the available worldwide DVCS data using the \texttt{Gepard} software. We find a significant reduction in the uncertainties of all CFFs after incorporating pseudo-data from single and double polarization asymmetries at the EicC, with particularly strong improvements in the sea-quark region.
\end{abstract}

\maketitle

\section{Introduction} \label{sec:Intro}

The three-dimensional distributions of quarks and gluons inside hadrons, encoded in terms of Generalized Parton Distributions (GPDs), can be probed through hard exclusive processes~\cite{Diehl:2003ny,Belitsky:2005qn,Guidal:2013rya,Diehl:2023nmm}.
Among these, the cleanest channel is Deeply Virtual Compton Scattering (DVCS)~\cite{Mueller:1998fv,Ji:1996nm,Ji:1996ek,Radyushkin:1997ki,Diehl:2005pc}, which is a central physics objective of future electron-ion colliders, where it will be explored across complementary kinematic regions with polarized beams.

The electron-ion collider (EIC) at Brookhaven National
Laboratory (BNL) will provide a high-precision experimental platform to study gluons in nucleons and nuclei~\cite{Accardi:2012qut,AbdulKhalek:2021gbh}.
Fixed-target experiments at Jefferson Lab (JLab), together with its potential upgrade~\cite{Accardi:2023chb},
investigate the valence-quark region and the onset of sea-quark contributions in great detail.
Strategically designed to complement the EIC and JLab programs, the Electron-Ion collider in China (EicC) aims to survey the sea-quark regime with unprecedented precision~\cite{CAO:2020EicC,CAO:2020Sci,Anderle:2021wcy}.
Alongside these efforts, the planned Large Hadron-Electron Collider (LHeC) at CERN would extend coverage deep into the small-$x$ domain~\cite{LHeCStudyGroup:2012zhm,LHeC:2020van}. Together, these facilities form a global program spanning a remarkably broad kinematic range. Such coverage will make it possible to reconstruct GPDs with controlled systematic uncertainties, leading to a more precise understanding of nucleon tomography~\cite{Ralston:2001xs,Diehl:2002he,Burkardt:2000za,Burkardt:2002hr}, spin structure~\cite{Ji:1996ek} and mechanical properties~\cite{Polyakov:2002yz} --- aspects of the nucleon's internal structure that currently remain insufficiently understood.

Although DVCS (and other exclusive processes involving GPDs) is sensitive to GPDs, it does not probe them directly.
Instead, GPDs enter observables through the Compton Form Factors (CFFs), which are integrals of the GPDs weighted by a hard-scattering kernel calculable order by order in perturbative  QCD.
Extracting the GPDs themselves therefore requires deconvoluting this highly nontrivial
relation --- a challenge associated
with the issue of ``shadow GPDs''~\cite{Bertone:2021yyz,Moffat:2023svr}.
The extraction and separation of individual CFFs from data are greatly facilitated by measured angular distributions and spin asymmetries, and constitute the first step toward determining GPDs from their associated form factors.
After several early attempts at local CFF extraction from single kinematic points~\cite{Moutarde:2009fg,Boer:2014kya,Kumericki:2013br},
global fits to the available data from JLab Hall A, CLAS, COMPASS, and HERMES collaborations have provided quantitative information on the leading-twist chiral-even CFFs.
This procedure is inherently complex due to the large number of unknown functions and the multiple independent kinematic variables on which they depend.
Non-parametric neural network (NN)~\cite{Forte:2002fg}, Gaussian Process regression \cite{Wang:2025obm}, and other flexible parametrization methods~\cite{Ye:2017gyb,Kotz:2023pbu} have proved useful for one- and two-dimensional problems such as standard Parton Distribution Functions (PDFs).
To minimize model dependencies and uncertainties arising from extrapolation and interpolation in multidimensional kinematic space, NN parametrizations of the CFFs offer the unique advantage of flexibility and hyperparameter choices~\cite{Kumericki:2011rz,Moutarde:2019tqa,CaleroDiaz:2025luc}.
Such parametrizations also enabled progress in the extraction of gravitational form factors from the DVCS data \cite{Kumericki:2019ddg} and in the flavor separation of key CFFs~\cite{Cuic:2020iwt,CLAS:2024qhy}.

Several categories of GPD models have been proposed, including the double distribution ansatz \cite{Musatov:1999xp}, representations in the conformal-moment space \cite{Mueller:2005ed}, string-inspired parametrization in AdS space~\cite{Mamo:2024jwp}, basis light-front quantization \cite{Xu:2021wwj}, and neural-networks approaches~\cite{Dutrieux:2021wll,Dotson:2025omi}.
Some of these have already been used in global fits to DVCS data~\cite{Kumericki:2007sa,Kumericki:2009uq,Moutarde:2018kwr,Cuic:2023mki,Guo:2023ahv,Guo:2025muf}.
However, the systematic uncertainties due to the model rigidity are not fully understood.
Upcoming efforts, aided by lattice-QCD calculations of GPDs~\cite{Constantinou:2020hdm,Lin:2020rxa,Holligan:2023jqh,Bhattacharya:2024qpp,Constantinou:2020hdm,Riberdy:2023awf}, within either the large-momentum effective theory~\cite{Ji:2013dva,Ji:2014gla} or the pseudo-distribution approach~\cite{Radyushkin:2017cyf}, are expected to yield more reliable uncertainty estimates.

At the current stage, quantifying the impact of new or proposed DVCS experiments on CFF extractions can be carried out efficiently using suitable NN architectures with the Monto Carlo replica method,
possibly combined with the  Bayesian reweighting procedure \cite{Moutarde:2019tqa}.
The relevance of the transversely polarized proton beam-spin asymmetries at EicC and their sensitivity to the corresponding dominant CFF have been demonstrated in \cite{Anderle:2021wcy} by such an approach.
There, the pseudo-data of DVCS measurements at EicC have been generated without considering the detector efficiency \cite{Cao:2023wyz}. 
In this work, we extend these previous studies by incorporating realistic detector coverage and efficiency, and by generating all measurable spin asymmetries.
Neural-network parametrization of the CFFs are then used to assess the impact of future EicC DVCS measurements on the determination of all leading-order CFFs.
Our analysis framework is built upon \texttt{Gepard}, a Python software package for the study of GPDs \cite{Kumericki:2007sa,Cuic:2023mki}.


\section{Simulation of DVCS and extraction of CFFs} 
\label{sec:DVCS}


\begin{figure}[htbp]
  \centering
  \includegraphics[width=0.95\textwidth]{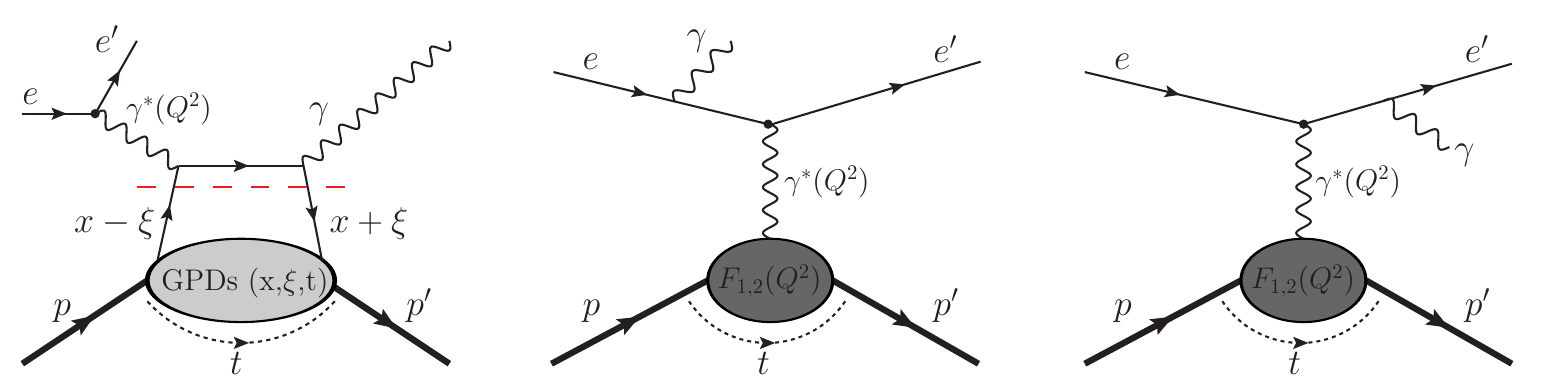}
  \caption{The deeply virtual Compton scattering (left) and the accompanying Bethe-Heitler process (middle and right).
  The $Q^2 = -q^2$ is the negative four-momentum squared of the virtual photon.
  Other kinematic variables are: $x$, the average longitudinal momentum fraction of the active quark; $\xi$, the half longitudinal momentum fraction transferred to the nucleon; and $t$, the squared four-momentum transferred to the nucleon.
  The $F_{1,2}$ are elastic form factors in the Dirac-Pauli representation. }
  \label{fig:DVCS}
\end{figure}

The exclusive photon electroproduction process $e p \to e' p' \gamma$ is the coherent sum of the electromagnetic Bethe-Heitler (BH) amplitude and the DVCS amplitude, facilitating access to GPDs, as shown in Fig. \ref{fig:DVCS}.
At leading twist the quark-helicity-conserving GPDs $H$, $E$, $\widetilde{H}$, and $\widetilde{E}$ contribute to DVCS through the corresponding CFFs $\mch$, $\mce$, $\widetilde{\mch}$ and $\widetilde{\mce}$.
The five-fold differential cross section is
\be
\difd\sigma(\phi,\phi_S) \equiv \frac{\difd\sigma^{e p \to e' p' \gamma}}{\difd \xbj \difd Q^2 \difd |t| \difd\phi \difd\phi_S} = \difd\sigma_{\textrm{UU}} [1 + h_l A_{\textrm{LU}} +h_p A_{\textrm{UL}} +h_l h_p A_{\textrm{LL}} + S_\textrm{T} ( A_{\textrm{UT}} + h_l A_{\textrm{LT}})]\,,
\ee
where $h_l/2$ ($h_p/2$) is the electron (proton) beam helicity, and $S_\textrm{T}$ is the component of the incoming proton spin transverse to the to virtual photon direction.
The kinematic variables are explained in the caption of Fig. \ref{fig:DVCS}, and at leading twist the skewness $\xi$ is related to Bjorken-$x$ by $\xi = x_B/(2-x_B)$.
Here $\phi$ is the azimuthal angle between the lepton plane and the real-photon production plane \cite{Bacchetta:2004jz}. 
The angle $\phi_S$ denotes the azimuthal orientation of the transverse proton-spin component with respect to the lepton scattering plane.
Integrating over $\phi_S$ reduces the cross section $\difd\sigma(\phi,\phi_S)$ to the four-fold one $\difd\sigma(\phi) \equiv {\difd\sigma^{e p \to e' p' \gamma}}/{\difd \xbj \difd Q^2 \difd |t| \difd\phi }$, in which the $\cos{\phi}$ modulation of the BH-DVCS interference term is sensitive to the real part of the CFFs ${\mathcal{H}}$ in case of proton DVCS:
\be \label{eq:xmodul}
\sigma^{\cos{\phi}}_{\textrm{UU},I} \propto \mbox{Re}\, \Big[{F_1 \mathcal{H}} +\xi (F_1+F_2)\mathcal{\widetilde{H}} -\frac{t}{4M^2} { F_2 \mathcal{E}}\Big] \,.
\ee

The longitudinally polarized proton or electron beam single-spin asymmetries are defined as
\bea \label{eq:asylong}
\textrm{A}_{\textrm{LU}} &=& \frac{\difd\sigma(\phi)^\rightarrow-\difd\sigma(\phi)^\leftarrow}{\difd\sigma(\phi)^\rightarrow+\difd\sigma(\phi)^\leftarrow} \,, \quad
\textrm{A}_{\textrm{UL}} = \frac{\difd\sigma(\phi)^\Rightarrow-\difd\sigma(\phi)^\Leftarrow}{\difd\sigma(\phi)^\Rightarrow+\difd\sigma(\phi)^\Leftarrow} \,,
\eea
where the arrows $\rightarrow$ ($\leftarrow$) or $\Rightarrow$ ($\Leftarrow$) refer to the electron or proton polarization parallel (anti-parallel) to the beam momentum.
The transverse proton beam-spin asymmetry is
\bea \label{eq:asytrans}
\textrm{A}_{\textrm{UT}} &=& \frac{\difd\sigma(\phi,\phi_S)-\difd\sigma(\phi,\phi_S+\pi)}{\difd\sigma(\phi,\phi_S)+\difd\sigma(\phi,\phi_S+\pi)}  \,. \quad
\eea
The double-spin asymmetries for a longitudinally polarized electron beam are
\bea
\textrm{A}_{\textrm{LL}} &=&  \frac{[\difd\sigma(\phi)^{\rightarrow\Rightarrow} + \difd\sigma(\phi)^{\leftarrow\Leftarrow}] - [(\difd\sigma(\phi)^{\leftarrow\Rightarrow}+\difd\sigma(\phi)^{\rightarrow\Leftarrow}]}{\difd\sigma(\phi)^{\rightarrow\Rightarrow} + \difd\sigma(\phi)^{\leftarrow\Leftarrow} + (\difd\sigma(\phi)^{\leftarrow\Rightarrow}+\difd\sigma(\phi)^{\rightarrow\Leftarrow}]} \,,  \\
\textrm{A}_{\textrm{LT}} &=& \frac{[\difd\sigma(\phi,\phi_S)^\rightarrow+\difd\sigma(\phi,\phi_S+\pi)^\leftarrow]-[\difd\sigma(\phi,\phi_S)^\leftarrow+\difd\sigma(\phi,\phi_S+\pi)^\rightarrow]}{\difd\sigma(\phi,\phi_S)^\leftarrow+\difd\sigma(\phi,\phi_S)^\rightarrow+\difd\sigma(\phi,\phi_S+\pi)^\rightarrow+\difd\sigma(\phi,\phi_S+\pi)^\leftarrow} \,,
\eea
for longitudinally and transversely polarized proton beam, respectively.
Depending on whether the electron beam helicity or proton spin is flipped,
different CFF combinations enter the cross section differences.
The azimuthal modulations of asymmetries, dominated by the interference term  and thus mostly linear in the CFFs, yield~\cite{Belitsky:2001ns,Belitsky:2005qn,Diehl:2005pc,Belitsky:2010jw,Kroll:2012sm,dHose:2016mda}:
\bea
   \textrm{A}^{\sin{\phi}}_{\textrm{LU},I}&\propto& \mbox{Im}\, \Big[{F_1 \mathcal{H}}
                     +\xi (F_1+F_2)\mathcal{\widetilde H}
                    -\frac{t}{4m^2} { F_2 \mathcal{E}}\Big]\,,  \\
   \left\{
   \begin{matrix}
     \textrm{A}^{\sin{\phi}}_{\textrm{UL},I} & \\
     \textrm{A}_{\textrm{LL},I}^{\cos{\phi}} &
   \end{matrix}
   \right\}
   &\propto&
   \left\{
   \begin{matrix}
     \mbox{Im} & \\
     \mbox{Re} &
   \end{matrix}
   \right\}
   \Big[\xi(F_1+{ F_2})({ \mathcal{H}}
                  +\frac{\xi}{1+\xi}\mathcal{E}) + {F_1\widetilde{\mathcal{H}}}-
      \xi(\frac{\xi}{1+\xi}F_1+\frac{t}{4M^2}{ F_2}){ \widetilde{\mathcal{E}}}\Big]\,, \qquad  \\
   \left\{
   \begin{matrix}
     \textrm{A}_{\textrm{UT},I}^{\sin{(\phi-\phi_s)}\cos{\phi}} & \\
     \textrm{A}_{\textrm{LT},I}^{\sin{(\phi-\phi_s)}\sin{\phi}} &
   \end{matrix}
   \right\}
   &\propto&
   \left\{
   \begin{matrix}
     \mbox{Im} & \\
     \mbox{Re} &
   \end{matrix}
   \right\}
   \Big[\frac{t}{4M^2}\big({ (1-\xi) F_2\mathcal{H}}-\frac{1}{1+\xi}{F_1\mathcal{E}}\big)
     -\xi^2\big(F_1+\frac{t}{4M^2}F_2\big)\big(\mathcal{H}+\mathcal{E}\big)  \nn \\
   &&\qquad  +\xi^2\big(F_1+F_2)\big(\widetilde{\mathcal{H}}
              + \frac{t}{4M^2}\widetilde{\mathcal{E}}\big)\Big]\,,  \\
   \left\{
   \begin{matrix}
     \textrm{A}_{\textrm{UT},I}^{\cos{(\phi-\phi_s)}\sin{\phi}} & \\
     \textrm{A}_{\textrm{LT},I}^{\cos{(\phi-\phi_s)}\cos{\phi}} &
   \end{matrix}
   \right\}
   &\propto&
   \left\{
   \begin{matrix}
     \mbox{Im} & \\
     \mbox{Re} &
   \end{matrix}
   \right\}
   \big[ \frac{t}{4M^2} \big({ (1-\xi) F_2\widetilde{\mathcal{H}}}-{\frac{\xi}{1+\xi} (F_1 +\xi F_2) \widetilde{\mathcal{E}}} + \frac{\xi}{1+\xi} (F_1 + F_2) \mathcal{E} \big)  \nn \\
   &&\qquad+ \frac{\xi^2}{1+\xi} \big( (F_1 + F_2) \mathcal{H} - F_1 \widetilde{\mathcal{H}} \big) \big] \,,
\eea
where ``$\textmd{Im}$'' and ``$\textmd{Re}$'' project the imaginary and real part of CFFs, respectively.
These relations show that different observables have different sensitivity to a given CFF.
The imaginary parts of CFFs ${{\mathcal{H}}}$ and
${\widetilde{\mathcal{H}}}$ remarkably manifest themselves in the $\sin{\phi}$ modulations of
single spin assymetries $\textrm{A}_{\textrm{LU}}$ and $\textrm{A}_{\textrm{UL}}$, respectively.
The corresponding real parts dominate the $\cos{\phi}$ modulations of the unpolarized cross section and of the $\textrm{A}_{\textrm{LL}}$.
The CFF $\mathcal{E}$ is most effectively accessed in transverse proton-spin asymmetries through the $\sin{(\phi-\phi_s)}\cos{\phi}$ modulation of $\textrm{A}_{\textrm{UT}}$ and $\textrm{A}_{\textrm{LT}}$,
while the $\cos{(\phi-\phi_s)}\cos{\phi}$ modulation of these asymmetries shows
some sensitivity to ${\widetilde{\mathcal{E}}}$. 
Because the BH process dominates the total cross section, extracting the CFFs with high precision requires extensive measurements over a broad kinematic range, including accurate determinations of cross sections and spin asymmetries and especially their azimuthal modulations.

\begin{widetext}
\begin{sidewaysfigure}
  \begin{center}
  {\includegraphics[width=0.9\textheight]{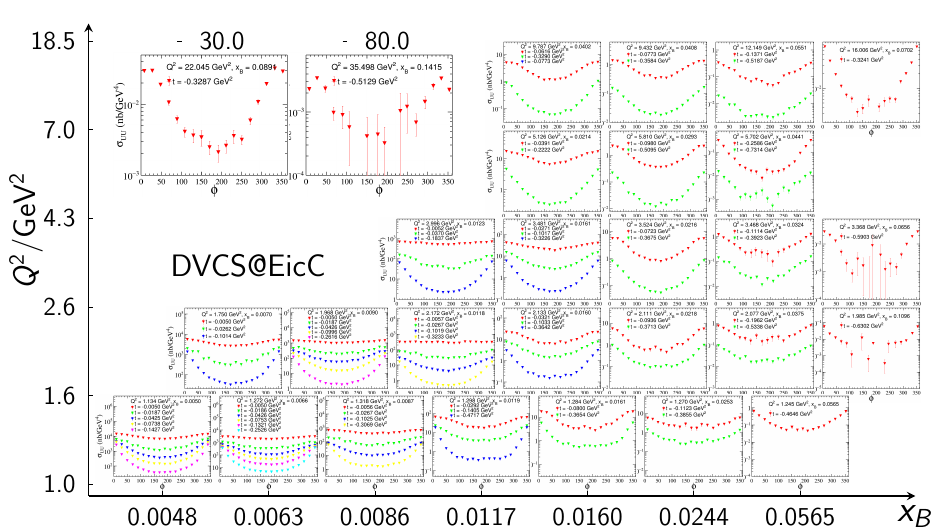}}
    \caption{The simulated four-fold unpolarized cross sections with statistic uncertainties in the bins of $\xbj$ and $Q^2$. In each panel different $t$ values in unit of GeV$^2$ are explicitly labeled. The systematic uncertainties are not included. The central values of pseudo-data are generated using the GK GPD model~\cite{Goloskokov:2009ia,Goloskokov:2011rd,Kroll:2012sm} and smeared by the statistic uncertainties.}
    \label{fig:EicC_pDVCS_proj}
  \end{center}
\end{sidewaysfigure}
\end{widetext}

The raw DVCS event samples at the EicC for a 3.5 GeV electron beam colliding with a 20 GeV proton beam are generated by the Monte Carlo (MC) generator MILOU~\cite{Perez:2004ig}, slightly modified from its original version~\cite{Aschenauer:2013hhw,Aschenauer:2017jsk} and temporarily neglecting the proton-dissociation background $ep \to e\gamma X$.
Previous study applied proper kinematic cuts to ensure DVCS dominance and removed resonance contributions by selecting events with photon-proton invariant mass above 2.0 GeV~\cite{Cao:2023wyz}.
In the present analysis, realistic detector acceptance and efficiency are included, assuming a 50 mrad beam crossing angle, following the EicC Conceptual Design Report (CDR) \cite{EicC:2025cdr}. 
A fast simulation framework incorporating acceptance, efficiency and resolution  --- previously established in EicC simulations of exclusive photo-production of heavy quarkonium \cite{Wang:2025obm,Wang:2023thy,Cao:2019gqo}, exotic hadron production \cite{Cao:2023rhu,Cao:2020cfx,Yang:2020eye,Cao:2019kst}, Sullivan process \cite{Lu:2025bnm}, and various inclusive processes \cite{Zeng:2022lbo} --- is employed.
The experimentally demanding yet essential recoil-proton detection is realized using optimized Roman pots inside the beampipe.
Two accelerator operation modes are considered: high-luminosity mode
$\mathcal{L} = 4 \times 10^{33}$ cm$^{-2}$ s$^{-1}$ with
Roman pots optimized for detection of particles with scattering angles larger than 10 mrad,
and complementary small-angular-divergence mode, allowing detection down to 5 mrad, with luminosity reduced to $1.1 \times 10^{33}$ cm$^{-2}$ s$^{-1}$.
Pseudo-data are generated assuming $3/4$ of running time in high-luminosity mode and $1/4$ in small-angular-divergence mode (see the \cite{EicC:2025cdr,He:2023svm} for details ).

The projected DVCS cross sections $\difd\sigma(\phi)$ assuming a year of data-taking are shown in Fig. \ref{fig:EicC_pDVCS_proj} together with their statistical uncertainties. They are organized into 69 kinematic bins in ($\xbj$, $Q^2$, $-t$), each further subdivided into 18 $\phi$-bins.
Exploiting the good detection efficiency it is feasible to measure the sea quark region  0.01 $< \xbj <$ 0.1 and the safely perturbative domain 2 GeV$^2 < Q^2 <$ 30 GeV$^2$.
Although the detection efficiency decreases at lower $\xbj$, it is compensated by the extremely high statistics achievable at $\xbj \sim 0.004$ when $Q^2$ approaches 1 GeV$^2$.
The detector coverage also provides welcome overlap with the JLab 12-GeV program in the valence region, but at generally higher $Q^2$ values at the EicC.
This safely perturbative domain without higher-twist corrections extend the high $Q^2$ lever arm for valence quark physics. Of particular interest is the largely unexplored domain $Q^2 \in$ [30.0, 80.0] GeV$^2$ with an average $-t = 0.51$ GeV$^2$, where sufficient statistics enable differential cross-section measurements.
The uncorrelated statistical uncertainties of the asymmetries in each bin are computed using a likelihood-based estimator,
\be \label{eq:staerror}
\delta\textrm{A} \propto \frac{1}{f_d P_\textrm{p} P_e} \sqrt{ \frac{1 - \textrm{A}}{N_{events}} } \,,
\ee
where $f_d$  denotes the dilution factor accounting for unpolarized background contributions, and the electron and proton beam polarizations are taken as $P_e =$ 80\% and $P_\textrm{p} = $ 70\%, respectively.
Here $N_{events}$ is the number of events obtained by scaling the generated cross sections to the assumed integrated luminosity of EicC. 

The absolute statistical {uncertainties} of the measured asymmetries {fall in the range of approximately 1\%–5\% across all considered kinematic
bins}.
The main sources of systematic uncertainty for the proposed measurement are radiative corrections and contamination from $\pi^0$ background channels.
Additional contributions to systematics include the beam-polarization uncertainties, acceptance corrections and lepton identification.
The aim of detector optimization is to control the total systematic uncertainty so that it remains comparable to,  or smaller than, the statistical uncertainty.

\begin{figure}[htbp]
    \centering
    \includegraphics[width=0.6\linewidth]{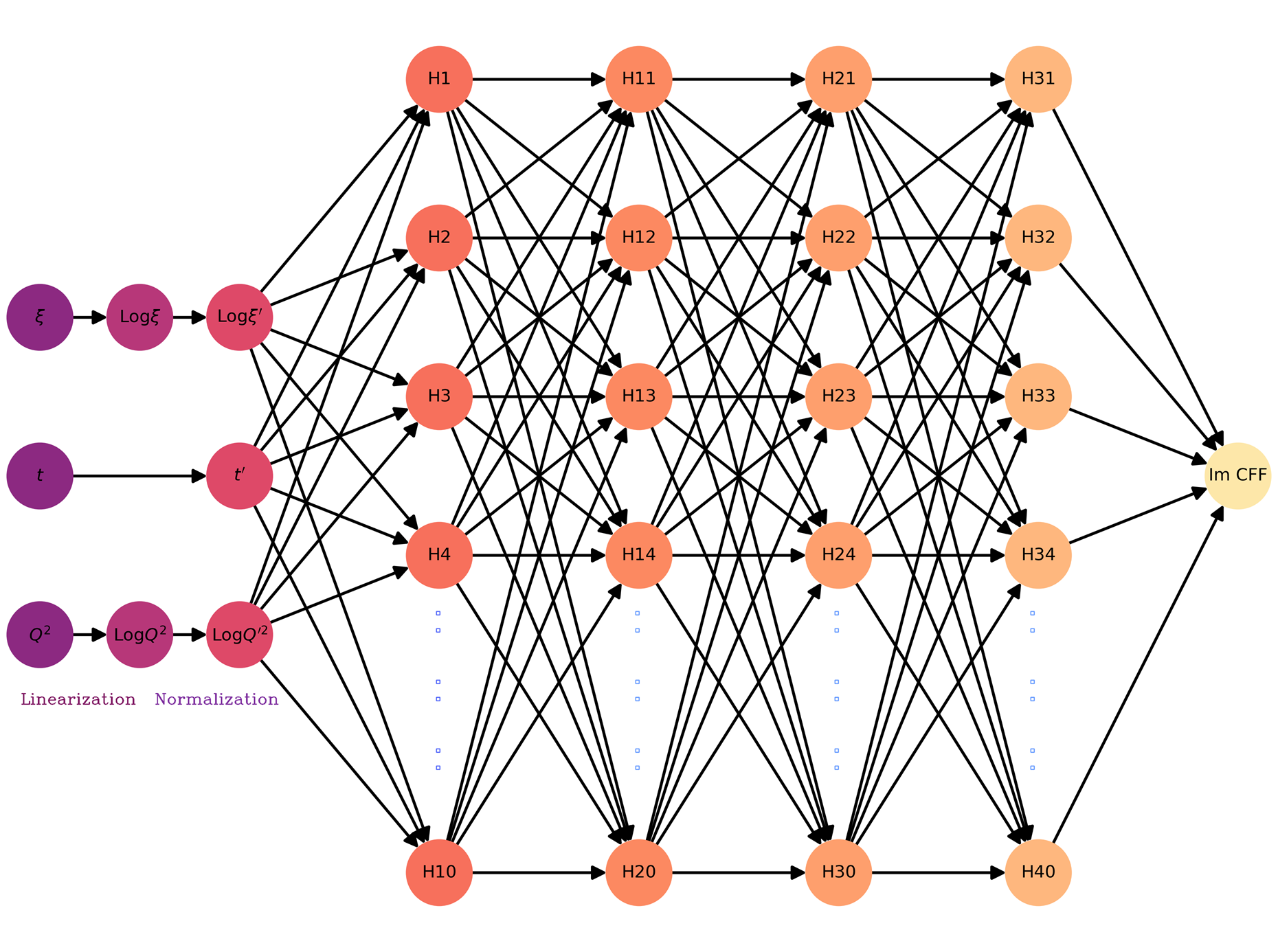}
    \caption{A single neural network parameterizes the imaginary part of one of the CFFs, while a similar net represents the real part. The input values ($\xi$, $t$, $Q^2$) after linearization and normalization are fed to three input neurons, and then propagated through the four hidden layers with ninety neurons each, so architecture is (3 $\to 90 \to 90 \to 90 \to 90 \to$ 1).}
    \label{fig:artNN}
\end{figure}

\begin{figure}[htbp]
  \begin{center}
  {\includegraphics*[width=0.6\textwidth]{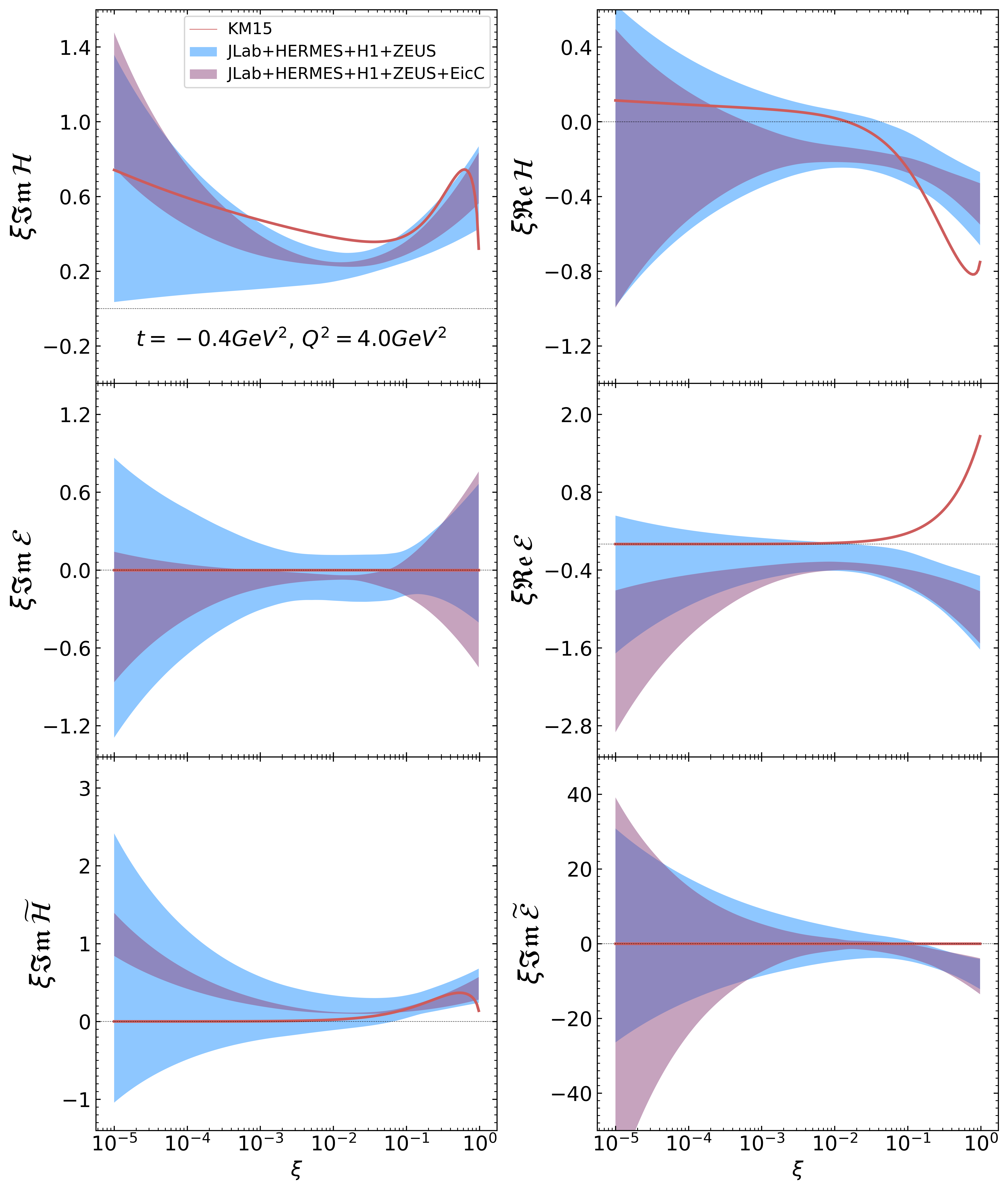}}
    \caption{The extraction of CFFs versus skewness $\xi$ at $Q^2 = $ 4.0 GeV$^2$ and $-t =$ 0.4 GeV$^2$.
    The blue and red error bands are uncertainties before and after including the pseudodata of asymmetries at EicC. } 
    \label{fig:CFFatEicCxi}
  \end{center}
\end{figure}
\begin{figure}[htbp]
  \begin{center}
  {\includegraphics*[width=0.6\textwidth]{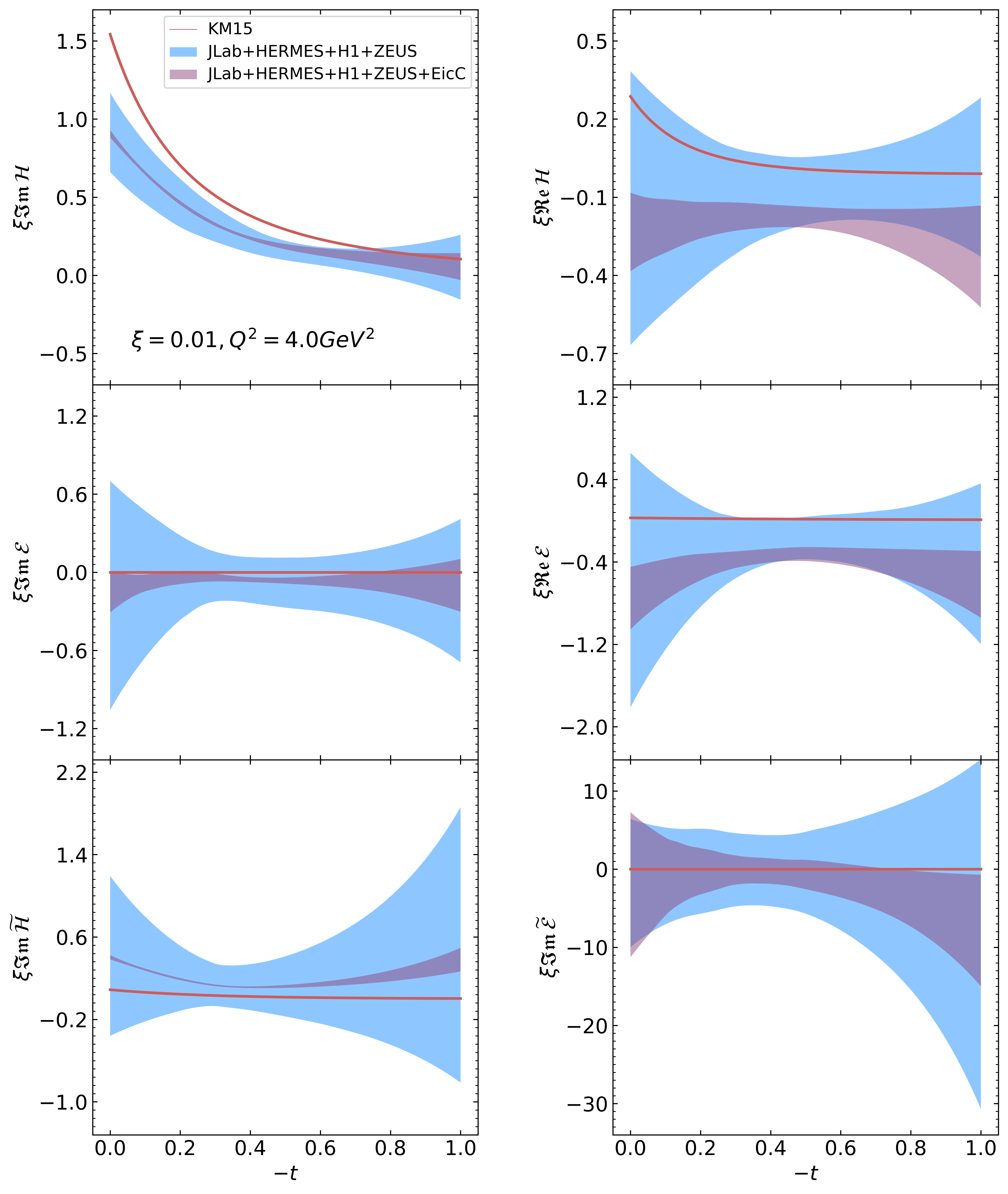}}
    \caption{The extraction of CFFs versus $-t$ at $Q^2 = $ 4.0 GeV$^2$ and $\xi =$ 0.01 GeV$^2$.
     }.
    \label{fig:CFFatEicCt}
  \end{center}
\end{figure}
\begin{figure}[htbp]
  \begin{center}
  {\includegraphics*[width=0.6\textwidth]{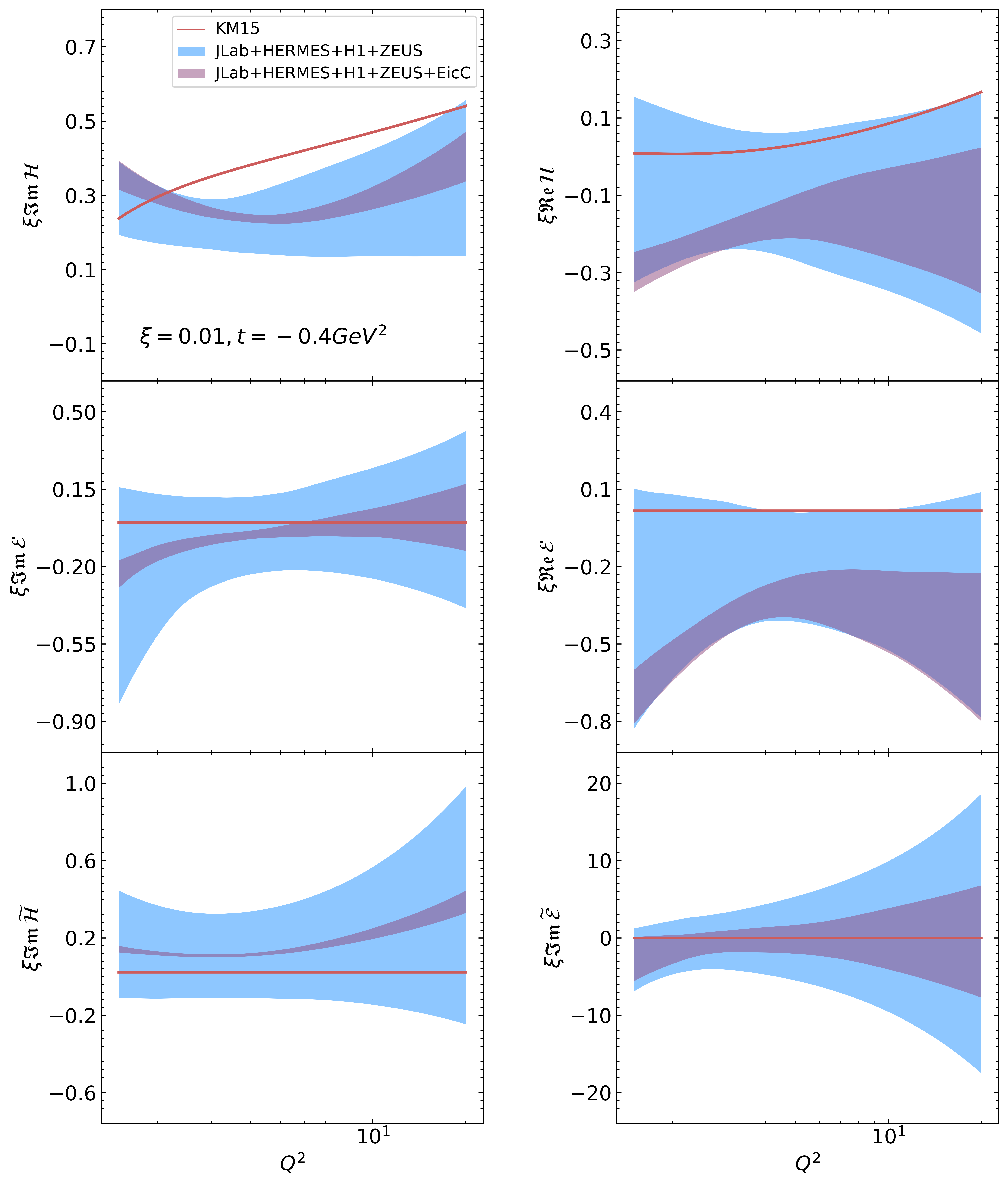}}
    \caption{The extraction of CFFs versus $Q^2$ at $\xi = $ 0.01 and $-t =$ 0.4 GeV$^2$.
     }.
    \label{fig:CFFatEicCQ2}
  \end{center}
\end{figure}

Progress in global fitting techniques --- particularly those representing the Compton form factors as neural networks --- has made it feasible to assess the impact of the above all-asymmetries pseudo-data on the CFFs extraction.
The connection between DVCS asymmetries and CFFs is constructed using the \texttt{Gepard} framework~\cite{Kumericki:2011rz,Kumericki:2019ddg,Cuic:2020iwt}.
Since current world data provide essentially no constraint on the real parts of $\widetilde{\mathcal E}$ and $\widetilde{\mathcal H}$, these are assumed to vanish in our analysis, following Ref.~\cite{Cuic:2020iwt}.
The NN architecture shown in Fig. \ref{fig:artNN} is implemented using the \texttt{PyTorch} machine-learning library, with the hyperparameters chosen for optimal extrapolation into unmeasured kinematic regimes.
To ensure the complete independence of the real and imaginary parts of CFFs, a modular neural network composed of independent {sub-networks} is constructed.
The activation function for neurons in the hidden layers is
\be \label{eq:activation}
    f(x) =
   \begin{cases}
     0 \, & \mbox{for } \, x < 0 \\
     e^x-1 \, & \mbox{for } \, x > 0
   \end{cases}
\ee
which provides the necessary non-linearity.
Training is performed via back-propagation algorithm using the Adam (Adaptive moment estimation) optimizer\cite{Kingma:2014vow}, minimizing an uncertainty-weighted Huber loss function \cite{Grigsby:2020auv,Almaeen:2024guo,Almaeen:2022imx,Adams:2024pxw,Hossen:2024qwo}.
The training data in the valence region include measurements from the JLab collaborations \cite{Defurne:2017paw,Defurne:2015kxq,dHose:2016mda,Hyde:2011ke,CLAS:2021gwi,Fucini:2021psq,Afanasev:2021twk},
COMPASS~\cite{COMPASS:2018pup} and HERMES~\cite{Airapetian:2012pg}, while the small-$\xbj$ region ($10^{-4} \lesssim \xbj \lesssim 10^{-3}$) is constrained by ZEUS and H1 data~\cite{Chekanov:2003ya,Chekanov:2008vy,Adloff:2001cn,Aktas:2005ty,Aaron:2009ac}.
All data sets are truncated with the kinematic cuts $Q^2 > 1.5~\text{GeV}^2,  -{t}/{Q^2} < 0.2$ with the purpose to suppress the higher-twist contribution.
The combined world dataset is randomly split into a $90\%$ training set and a $10\%$ test set and data leakage and overfitting are controlled using standard early-stopping technique.
Linearization and normalization of the input variables enable fast, high precision convergence, making it practical to generate a large number of Monte-Carlo replica datasets via bootstrapping in order to propagate statistical uncertainties from the data to the extracted CFFs.
The network depth, width, and choice of loss function are optimized based on fitting
and validation performance.

\begin{table}[htbp]
\caption{\label{tab:dataset}{The fitted $\chi^2$ per data point before and after including the EicC pseudo-data of asymmetries.}}
\begin{ruledtabular}
\begin{tabular}{l c c c c}
\textbf{Observable}	& \textbf{Collaboration}	& \textbf{Ref.}					& \textbf{$\chi^2$}	& \textbf{$\chi^2_{\textrm{w/ EicC}}$} \\
\hline
$\textrm{A}_{\textrm{LU}}$ & CLAS   & \cite{CLAS:2015bqi}  & 1.30						& 1.47 \\
$\textrm{A}_{\textrm{LU}}$ & HERMES & \cite{HERMES:2012gbh} & 0.95						& 0.81 \\
$\textrm{A}_{\textrm{UT}}$ & HERMES & \cite{HERMES:2008abz}  & 1.50						& 1.01 \\
$\textrm{A}_{\textrm{LL}}$ & CLAS   & \cite{CLAS:2015bqi}  & 2.09						& 2.12 \\
$\textrm{A}_{\textrm{UL}}$ & CLAS   & \cite{CLAS:2015bqi}  & 3.54						& 1.13 \\
$\sigma_{\textrm{UU}}$  & H1 & \cite{H1:2007vrx}  & 0.69						& 0.63 \\
$\sigma_{\textrm{UU}}$  & H1 & \cite{H1:2005gdw}  & 0.90						& 0.64 \\
$\sigma_{\textrm{UU}}$  & ZEUS  & \cite{ZEUS:2008hcd}  & 0.98						& 1.05 \\
$\textrm{A}_{\textrm{LU}}$ & CLAS  & \cite{CLAS:2015uuo}  & 0.43						& 0.42 \\
$\textrm{A}_{\textrm{LU}}$ & HALL A  & \cite{JeffersonLabHallA:2015dwe}  & 1.05						& 1.14 \\
$\sigma_{\textrm{UU}}$  & CLAS    & \cite{CLAS:2015uuo}  & 0.95						& 0.88 \\
$\sigma_{\textrm{UU}}$ & HALL A  & \cite{JeffersonLabHallA:2015dwe}  & 1.64						& 1.80 \\
$\sigma_{\textrm{UU}}$ & HALL A  & \cite{Defurne:2017paw}  & 1.21						& 0.90 \\\hline
Total   &  &   & 1.1 & 0.9 \\
\end{tabular}
\end{ruledtabular}
\label{tab:dataset-summary}
\end{table}

The actual $\chi^2$ per data point is in the range of 1.0 $\sim$ 2.0 for each data set, see Table \ref{tab:dataset}.
The skewness $\xi$-dependence of the CFFs extracted from existing world data at $Q^2 = $ 4.0 GeV$^2$ and $-t =$ 0.4 GeV$^2$ is shown by the blue band in Fig. \ref{fig:CFFatEicCxi}. Similarly,
the $-t$ dependence  at $Q^2 = $ 4.0 GeV$^2$ and $\xi =$ 0.01 GeV$^2$ is illustrated in Fig. \ref{fig:CFFatEicCt}, while the $Q^2$ dependence at $\xi = 0.01$ and $-t =$ 0.4 GeV$^2$ is displayed in Fig. \ref{fig:CFFatEicCQ2}.
In the valence region, these results agree with conclusions of previous analyses.
The extrapolation behavior is consistent with that obtained using \textit{PARTONS} \cite{Moutarde:2019tqa}, though our uncertainties are relatively smaller, likely due to our assumption of vanishing real part of $\widetilde{\mathcal E}$ and $\widetilde{\mathcal H}$.

The central values of the CFFs,  as well as the predicted cross sections and asymmetries beyond the valence region are consistent with zero within large uncertainties, reflecting the marginal impact of current data in the sea-quark and gluon domains.
For this reason, the pseudo-data for unpolarized cross sections are not included in the present impact study, since their relative statistical uncertainties are inversely proportional to $N_{events}$.
Consequently, a substantial reduction in the uncertainty of $\mbox{Re} \mathcal{H}$ should not be expected after including the EicC pseudo-data, as suggested by Eq. \eqref{eq:xmodul}.
Within the NN framework, the predicted asymmetry magnitudes are smeared by randomly resampling the central values according to the generated statistical uncertainties of Eq. \eqref{eq:staerror}, and the networks are then retrained on the full set of measured and simulated experimental data to assess the genuine impact of the EicC machine.
Although realistic systematic uncertainties would enlarge the final error bands, the presented global analysis clearly demonstrates that the EicC can prominently improve the precision of all relevant CFFs down to $\xi \sim 10^{-4}$. In particular, the uncertainties of CFFs in sea-quark region are significantly reduced, as indicated by the light-red bands in Figs. \ref{fig:CFFatEicCxi}.
The $-t$ dependence of CFFs in the sea regime is also extracted with good precision up to 1.0 GeV$^2$ (Fig. \ref{fig:CFFatEicCt}); thus EicC will provide an accurate spatial tomography of the sea quarks inside nucleon through Fourier transform, --- albeit necessarily relying on assumptions or models to extrapolate beyond the experimentally accessible kinematic region.
Likewise, the influence of EicC data on the $Q^2$ dependence of CFFs is evident up to 20 GeV$^2$ in the sea domain (Fig. \ref{fig:CFFatEicCQ2}), opening the possibility of GPD deconvolution from DVCS data via QCD scale evolution across a broad $Q^2$ range.

On the methodological side, the EicC pseudo-data provide an important validation of the extraction strategy over a wide kinematic range.
The error bands of CFFs obtained after including the EicC pseudo-data remain consistent with those derived solely from existing measurements, as expected, which serves as a non-trivial closure test
of the NN procedure.
However, noticable deviation exceeding 1$\sigma$ appear in the leading CFFs $\mbox{Re} \mathcal{H}$ and $\mbox{Re} \mathcal{E}$ in certain regions, reflecting the strong clustering of current data in a limited portion of phase space. 
With more experimental input, improved lattice-QCD constraints, and refined theoretical guidance, neural-network methods offer promising solution to the challenge of reaching balance between bias minimization and flexibility of the model.

\section{Summary and Conclusion}

The EicC is designed to probe the sea-quark region, providing kinematic coverage
between JLab and EIC at BNL.
The DVCS data of unprecedented accuracy at EicC are expected to yield a considerable improvement in understanding of proton tomography in terms of GPD.
A previous Bayesian reweighting study \cite{Cao:2023wyz} showed that even a single day of data taking with a transversely polarized proton beam at the EicC would already surpass the constraining
power of existing HERMES fixed-target measurements in the
sea-quark regime~\cite{Airapetian:2008aa}.

Neural-network parameterizations of CFFs enable
a precise extraction and disentanglement of the CFFs with a true global fit over all observables.
In this work, we constructed a flexible NN architecture to simultaneously extract the CFFs from existing data and to assess the constraining power of future EicC asymmetry measurements.
A significant reduction in the uncertainties of all relevant CFFs is obtained when including the projected EicC data.
This also constitutes a valuable closure test of the NN methodology, pointing toward the need for additional theoretical input to further stabilize extrapolations.
The framework is readily extendable to incorporate forthcoming datasets, including those expected from JLab~\cite{JeffersonLabSoLID:2022iod} and the EIC~\cite{Anderle:2021wcy,Aschenauer:2025cdq}.
It is feasible to incorporate higher-twist contributions into our global fit in the near future \cite{Braun:2025xlp}.
Only then would the importance of the extension to high $Q^2$ domain become clear for valence quark physics through a more direct comparison between EicC and JLab.

The type of work presented here represents a step toward reliable extraction of GPDs at LO by means of flexible GPD parametrizations in the same manner as PDFs are extracted from multi-dimensional structure functions.
However, achieving global fits using NLO \cite{Braun:2022qly,Braun:2025xlp} and NNLO \cite{Braun:2020yib,Braun:2021grd,Braun:2022bpn,Ji:2023xzk} framework remains a major challenge \cite{Cuic:2023mki}.






\bigskip

\begin{acknowledgments}

We are grateful to Aiqiang Guo, Yu-Tie Liang, and Weizhi Xiong for providing us the fast simulation of the EicC detector. This work is supported by the National Key R\&D Program of China under Grant No. 2023YFA1606703, the National Natural Science Foundation of China  (Grants Nos. 12547111, 12235008, and 12405105), and
Grant No. PK.1.1.10.0004 co-financed by the European Union and through the European Regional Development Fund - Competitiveness and Cohesion Programme 2021-2027.
TF also thanks the Hebei Natural Science Foundation with Grant Nos. A2022201017 and A2023201041, and Natural Science Foundation of Guangxi Autonomous Region with Grant No. 2022GXNSFDA035068.
We gratefully acknowledge the support of High Performance Computing Cluster of the Southern Nuclear Science Computing Center(SNSC).

\end{acknowledgments}

\bigskip

\bibliography{ref_CFFNNatEicC.bib}

\end{document}